\documentclass[12pt]{iopart}
\usepackage{iopams}
\begin{document}
\jl{1}

\title[Remarks on quantization of Pais-Uhlenbeck oscillators]%
{Remarks on quantization of Pais-Uhlenbeck oscillators}
\author{E.V.Damaskinsky\dag\quad and\quad M.A.Sokolov\ddag}

\address{\dag\
Department of Mathematics, Technical University of Defence Constructing,
Zacharievskaya 22, Saint-Petersburg, Russia. e-mail  evd@pdmi.ras.ru}
\address{\ddag\
Department of Physics, Saint-Petersburg Institute of Mashine Building,
Poliustrovskii pr-t 14, 195197, Saint-Petersburg, Russia.
e-mail  mas@MS3450.spb.edu}

\begin{abstract}
This work is concerned with a quantization of the Pais-Uhlenbeck
oscillators from the point of view of their multi-Hamiltonian structures.
It is shown that the $2n$-th order oscillator with a simple spectrum
 is equivalent to the usual anisotropic $n$ - dimensional oscillator.
\end{abstract}

\vspace{1.5cm}
\textbf{1.}
This work is concerned with a quantization of the Pais-Uhlenbeck
oscillators from the point of view of their multi-Hamiltonian structures.
The family of such oscillators was introduced in the paper
\cite{PU} as a toy model to study field theories with higher derivative
terms. Evolution of the Pais-Uhlenbeck $2n$-th order oscillator is defined
by the following equation

\begin{equation}
\label{no}
\prod _{i=1}^n(\frac{\mathrm{d}^2}{\mathrm{d}t^2}+{\omega }_i^2)x = 0
\end{equation}
where $\Omega =({\omega }_i),\,i = 1, ... ,n$  is a set of positive
parameters (frequencies).
Equation (\ref{no}) can be obtained by variation of the Lagrangian

\begin{equation}
\label{no-lagr}
L=-x\left(\prod _{i=1}^n(\frac{\mathrm{d}^2}{\mathrm{d}t^2}+
{\omega }_i^2)\right)x.
\end{equation}
Introducing the natural oscillator coordinates
\begin{equation}
\label{no-coord}
q_k=
\prod _{i=1}^{k-1}(\frac{\mathrm{d}^2}{\mathrm{d}t^2}+{\omega }_i^2)
\hspace{-2mm}
\prod _{i=k+1}^{n}(\frac{\mathrm{d}^2}{\mathrm{d}t^2}+{\omega }_i^2)x
\end{equation}
and conjugate momenta Pais and Uhlenbeck proved that the Hamiltonian
related to the Lagrangian (\ref{no-lagr}) has the form

\begin{equation}
\label{no-ham}
H_{PU} = \frac{1}{2}\sum _i^n(-1)^{i-1}(p_i^2+{\omega }_i^2q_i^2).
\end{equation}
It follows from this expression that the Hamiltonian is not positive
defined. Recently, the quantization of the fourth order Pais-Uhlenbeck
oscillator was curried out in some details by Mannheim and Davidson
\cite{ManDav} (see \cite{Smilga}, as well).
The starting point for the authors of \cite{ManDav} was the Lagrangian
(\ref{no-lagr}) with $n=2$. Using the Dirac method they obtained the
Hamiltonian (see below first expression from (\ref{fo-int}))
which is different in its form from the Pais-Uhlenbeck one (\ref{no-ham}).
But the quantization of this new Hamiltonian leads to the
Hilbert space $\cal H$ containing negative norm states
(this is because of improper sign of the commutator
of a pair of creation and annihilation operators).
This result is treated usually as the intrinsic property of
theories with higher derivative terms.

However, there are arguments that a satisfactory quantization of the
Pais-Uhlenbeck oscillators can be curried out. For simplicity, let us
consider the fourth order oscillator. In terms of the oscillator
coordinates (\ref{no-coord})
\[
q_1 = \frac{\mathrm{d}^2x}{\mathrm{d}t^2}+{\omega }_2^2x,\quad
q_2 = \frac{\mathrm{d}^2x}{\mathrm{d}t^2}+{\omega }_1^2x,
\]
and momenta (or velocities, which are identical to the momenta in this case)
\[
p_1 = \frac{\mathrm{d}q_1}{\mathrm{d}t},\quad
p_2 = \frac{\mathrm{d}q_2}{\mathrm{d}t}
\]
equation (\ref{no}) can be rewrited equivalently as a canonical system of
Hamiltonian equations of motion for the two-dimensional anisotropic
oscillator

\begin{equation}
\label{fo-s-canon}
\begin{tabular}{l}
$
\frac{\mathrm{d}q_1}{\mathrm{d}t} =
\frac{\partial H_C}{\partial p_1} = p_1, \quad
\frac{\mathrm{d}p_1}{\mathrm{d}t} =
-\frac{\partial H_C}{\partial q_1} = -{\omega }_1^2q_1,
$\\[12pt]
$
\frac{\mathrm{d}q_2}{\mathrm{d}t} =
\frac{\partial H_C}{\partial p_2} = p_2, \quad
\frac{\mathrm{d}p_2}{\mathrm{d}t} = -\frac{\partial H_C}{\partial p_2} =
-{\omega }_1^2q_2
$
\end{tabular}
\end{equation}
where
\begin{equation}
\label{fo-canon-ham}
H_C = \frac{1}{2}(p_1^2+{\omega }_1^2q_1^2)+
    \frac{1}{2}(p_2^2+{\omega }_2^2q_2^2).
\end{equation}
The canonical quantization of this oscillator leads to the usual
commutation relations among the creation and annihilation operators and
to the standard Hilbert space with a positive norm. Thus, it is seems
natural to deduce a relevant (for a quantization ) Hamiltonian
formulation of the model directly from  the equation of motion (\ref{no})
omitting a Lagrangian formulation.

The difference of the Hamiltonians (\ref{no-ham}) (for $n = 2$)
and (\ref{fo-canon-ham}) tells us that the Pais-Uhlenbeck
oscillator equation of motion can be obtained using nonequivalent
Hamiltonian structures. In another words this oscillator is a
bi-Hamiltonian system \cite{M}, \cite{KR}. In the case of
the classical fourth order oscillator  this fact was established
in \cite{BK}.

In the present paper we shall consider the quantization of the
Pais-Uhlenbeck oscillators from the point of view of their
multi-Hamiltonian nature.
It will be shown that the $2n$-th order oscillator with a simple spectrum
(all frequencies from the set $\Omega$ are different)
 is equivalent to the usual anisotropic $n$ - dimensional oscillator.
We shall start from the case of the fourth order oscillator and then
the related formulas for the general case will be written.

\textbf{2.}
The equation of motion of the forth order Pais-Uhlenbeck  oscillator
($n = 2$ in (\ref{no})) has the form

\begin{equation}
\label{fo}
\frac{\mathrm{d}^4x}{\mathrm{d}t^4}
+ ({\omega }_1^2+{\omega }_2^2)\frac{\mathrm{d}^2x}{\mathrm{d}t^2}
+ {\omega }_1^2{\omega }_2^2x=0.
\end{equation}
It can be written in the form of a system of first order equations

\begin{equation}
\label{fo-s}
\frac{\mathrm{d}x_1}{\mathrm{d}t} = x_2, \quad
\frac{\mathrm{d}x_2}{\mathrm{d}t} = x_3, \quad
\frac{\mathrm{d}x_3}{\mathrm{d}t} = x_4, \quad
\frac{\mathrm{d}x_4}{\mathrm{d}t} =
-({\omega}_1^2 +{\omega}_2^2)x_3 - {\omega}_1^2 {\omega}_2^2x_1,
\end{equation}
where $x_i,\,i=1, ... ,4,$ are local coordinates of the "phase"
space ($x_1=x$ is the coordinate in the original space,
$x_2$ is the velocity and so on).
Integral curves of the vector field

\begin{equation}
\label{fo-vf}
\mathbf{V}=
x_2\partial _1+x_3\partial _2+x_4\partial _3-(({\omega }_1^2
+ {\omega }_2^2)x_3+{\omega }_1^2{\omega }_2^2x_1)\partial _4,
\quad
\partial _i = \frac{\partial }{\partial x_i},
\end{equation}
are exactly solutions of the system (\ref{fo-s})
(see, for instance, \cite{A},\cite{O}).
The simplest way to obtain integrals of motion of the
oscillator is to use the following equation

\begin{equation}
\label{fo-const}
\mathbf{V}(H)= 0.
\end{equation}
Note here that for this purpose in the paper \cite{BK} was applied
the known solution of Eq. (\ref{fo}). In the local coordinates
Eq. (\ref{fo-const}) has the form
\[
x_2\partial _1H+x_3\partial _2H+x_4\partial _3H
-(({\omega}_1^2+{\omega}_2^2)x_3
+{\omega}_1^2{\omega }_2^2x_1)\partial _4H = 0.
\]
Since the components of the field $\mathbf{V}$ are homogeneous linear
coordinate functions then the analytic solutions of Eq. (\ref{fo-const})
are homogeneous polynomials in $x_i$.
For the considered Pais-Uhlenbeck oscillator we have two independent
positive defined quadratic integrals of motion

\begin{equation}
\label{fo-int+}
\begin{tabular}{l}
$H_1=\frac 12(x_4+{\omega }_2^2x_2)^2
+\frac 12{\omega }_1^2(x_3+\omega _2^2x_1)^2,
\quad $ \\[8pt]
$H_2=\frac 12(x_4+{\omega }_1^2x_2)^2
+\frac 12{\omega }_2^2(x_3+\omega _1^2x_1)^2.$
\end{tabular}
\end{equation}
Taking into account the above definition of the oscillator coordinates
$q_i$ and related momenta $p_i$ one can see that the Hamiltonian
(\ref{fo-canon-ham}) is the sum of integrals $H_1$ and $H_2$
from (\ref{fo-int+})

\[H_C = H_1 +H_2,\]
whereas the Pais-Uhlenbeck Hamiltonian is their difference
\[H_{PU} = H_1  - H_2.\]
Let us point out
that the coordinates $q_1,$  $q_2$ and the integrals
(\ref{fo-int+}) are degenerate in the case
${\omega }_1 = {\omega }_2.$
The simplest linear combinations  of $ H_1$ and $H_2$

\begin{equation}
\label{fo-int}
\begin{tabular}{l}
$C_1= \frac{{\omega }_1^2H_1-{\omega }_2^2H_2}
{{\omega }_1^2-{\omega }_2^2} =
-\frac 12{\omega }_1^2{\omega }_2^2x_2^2
+\frac 12({\omega }_1^2+{\omega }_2^2)x_3^2 +\frac 12x_4^2
+{\omega }_1^2{\omega }_2^2x_1x_3, \quad $
\\[8pt]
$C_2= -\frac{H_1-H_2}{{\omega }_1^2-{\omega }_2^2} =
\frac 12{\omega }_1^2{\omega }_2^2x_1^2
+\frac 12({\omega }_1^2+{\omega }_2^2)x_2^2 -\frac 12x_3^2 +x_2x_4,$
\end{tabular}
\end{equation}
gives us the pair of integrals which are distinct at
${\omega }_1 = {\omega }_2$, but $C_1$ and $C_2$ are not positive
defined.  The integral $C_1$ was obtained in
\cite{ManDav} and it was taken as a Hamiltonian of the fourth order
Pais-Uhlenbeck oscillator.  The notation of
\cite{ManDav} and ours are related by the following formulas

\[x_1 = p_q,\quad
x_2 = \gamma{\omega }_1^2{\omega }_2^2q,\quad
x_3 = \gamma{\omega }_1^2{\omega }_2^2x,\quad
x_4 = {\omega }_1^2{\omega }_2^2p_x.
\]

Now introduce a Poisson structure for the considered oscillator.
Recall that a Poisson structure on a manifold is defined by a
rank two contravariant tensor field $\Pi$ which is skew-symmetric
\[ \Pi ^{ij} = - \Pi ^{ji}\]
and satisfies the condition
\[
[[\Pi,\Pi]]^{ijk} :=
\sum _m (\Pi ^{mi}\partial _m\Pi ^{jk}+
\Pi ^{mj}\partial _m\Pi ^{ki}+\Pi ^{mk}\partial _m\Pi ^{ij}) = 0.
\]

Any Poisson structure induces a Poisson brackets $\{F,G\}$
of arbitrary differentiable functions $F,G$ on a manifold $M$.
In local coordinates $x_i$ the brackets are defined by the
formula

\[
\{F,G\} = \Pi ^{ij}\partial _iF\partial _jG.
\]

The vector field $\mathbf{V}$ is called locally Hamiltonian if
there is a Poisson structure $\Pi$ such that

\begin{equation}
\label{fo-lieder}
\mathcal{L}_{\mathbf{V}}(\Pi)=0
\end{equation}
where $\mathcal{L}_{\mathbf{V}}$ defines the Lie derivative along
the field $\mathbf{V}$. In local coordinates the above relation
has the form

\[
V^k\frac{\partial \Pi ^{ij}}{\partial x^k}
-\frac{\partial V^i}{\partial x^k}\Pi ^{kj}
-\Pi ^{ik}\frac{\partial V^j}{\partial x^k}=0,
\]
where $V^i$ and $\Pi ^{ij}$ are components of the vector field
$\mathbf{V}$ and the Poisson tensor $\Pi$. If there is such a
differentiable function $H$ that ${\mathbf{V}}$ can be represented in
the form

\[
\mathbf{V}_H(.) = \{.,H\},
\]
then $H$ is called a Hamiltonian. In this case equations of motion take
the canonical form

\begin{equation}
\label{fo-hamilt}
\frac{\mathrm{d}x_i}{\mathrm{d}t}=\{x_i,H\}.
\end{equation}

Considering the relation (\ref{fo-lieder}) as an equation one can
obtain a Poisson tensor $\Pi$ related to the field ${\mathbf{V}}$.
The simplest solution of this equation is a two-parameter nondegenerate
Poisson tensor with constant components.
Its components are represented in the following
table  ($f$ and $g$ are arbitrary parameters)

\begin{equation}
\label{fo-poisconst}
\left[\Pi ^{ij}_{f,g}\right] =
\left[
\begin{tabular}{cccc}
$0$&$f$  & $0$ &$g$  \\
$-f$&$0$  &$-g$  &$0$  \\
$0$&$g$  &$0$  &
$ -{\omega }_1^2{\omega }_2^2f-({\omega }_1^2+{\omega }_2^2)g$ \\
$-g$&$0$  &
${\omega }_1^2{\omega }_2^2f+({\omega }_1^2+{\omega }_2^2)g$  &$0$
\end{tabular}
\right].
\end{equation}
This Poisson tensor $\Pi _{f,g}$ is obviously skew-symmetric
and satisfy the condition $[[\Pi,\Pi]]^{ijk}=0$ in view of its
constancy. It induces the following Poisson brackets for the coordinate
functions

\begin{equation}
\label{fo-brackconst}
\begin{tabular}{ll}
$\{x_1,x_2\}_{f,g}=f, \quad$ &
$\{x_1,x_4\}_{f,g}=g, \quad$ \\[8pt]
$\{x_2,x_3\}_{f,g}=-g,\quad$ &
$\{x_3,x_4\}_{f,g} =
-{\omega }_1^2{\omega }_2^2f-({\omega }_1^2+{\omega }_2^2)g.$
\end{tabular}
\end{equation}
It is not difficult to check that
the dynamical equations (\ref{fo-s}) are generated by these
brackets

\[
\frac{\mathrm{d}x_i}{\mathrm{d}t}=\{x_i,H\}_{f,g}
\]

together with the Hamiltonian function

\begin{equation}
\label{fo-ham+}
H=a_1H_1+a_2H_2,
\end{equation}
where the coeffitients $a_i$ have the form

\begin{equation}
\label{fo-ham-const}
a_1=\frac 1{({\omega }_2^2-{\omega }_1^2)({\omega }_2^2f+g)},
\qquad
a_2=-\frac 1{({\omega }_2^2-{\omega }_1^2)({\omega }_1^2f+g)},
\end{equation}
and can be choosing positive.
Thus the dynamical equations
(\ref{fo-s}) (and the field $\bf V$ itself) are Hamiltonian ones
and the two-parameter function $H$ plays the role of a Hamiltonian.
Remark that the integrals of motion $C_1$ and
$C_2$ are in involution in respect with these brackets

\[
\{C_1,C_2\}_{f,g}=0.
\]

In the classical case the parameters $f$ and $g$ can be taken
either arbitrary or fixed in any appropriate manner. For instance,
we can put

\[
f=-\frac 1{{\omega}_1^2{\omega }_2^2},\quad\,g=0.
\]

This choice gives the following nonzero components of the
Poisson tensor $\Pi _1$

\begin{equation}
\label{fo-pi1}
\{x_1,x_2\}_1=-\frac 1{{\omega}_1^2{\omega }_2^2},\quad
\{x_3,x_4\}_1 = 1.
\end{equation}
Other simple choice

\[f=0,\quad\,g=1,\]
gives
\begin{equation}
\label{fo-pi2}
\{x_1,x_4\}_2=1,\quad
\{x_2,x_3\}_2=-1,\quad
\{x_3,x_4\}_2 = -({\omega }_1^2+{\omega }_2^2)
\end{equation}
for the components of $\Pi _2$. Both mentioned Poisson
structures generate the dynamical equations (\ref{fo-s})

\begin{equation}
\label{fo-biham}
\frac{\mathrm{d}x_i}{\mathrm{d}t}=\{x_i,C_1\}_1 = \{x_i,C_2\}_2.
\end{equation}
Thus the fourth order Pais-Uhlenbeck oscillator is a bi-hamiltonian system
(\cite{M},\cite{KR},\cite{O}) with the Hamiltonians
$C_1,\,C_2$ and the Poisson structures $\Pi _1,\,\Pi _2.$
First from these structures has been used in \cite{ManDav}.
Let us note in conclusion, that there is no a constant Poisson
structure which generates the equations (\ref{fo-s}) together with
any of the Hamiltonians (\ref{fo-int+}).

\textbf{3.}
The quasiclassic quantization of the Poisson structure
(\ref{fo-poisconst}) can be considered as exact because of
the  dynamical equations (\ref {fo-s}) of the fourth order
Pais-Uhlenbeck oscillator are reducible to the canonically quantized
form (\ref{fo-s-canon}) by the linear transformation.
Assume that the hermitian operators $\hat x_i,\,i=1, ... ,4,$
related to the dynamical variables of the classical system $x_i,$
subject the following  commutation relations

\begin{equation}
\label{fo-q-commut}
\begin{tabular}{l}
$[\hat x_1,\hat x_2]_{f,g}=\mathrm{i}\hbar f,
\quad
[\hat x_1,\hat x_4]_{f,g}=\mathrm{i}\hbar g,$
\\[6pt]
$[\hat x_2,\hat x_3]_{f,g}=-\mathrm{i}\hbar g,
\quad
[\hat x_3,\hat x_4]_{f,g}=-\mathrm{i}\hbar {\omega }_1^2{\omega }_2^2f
-\mathrm{i}\hbar ({\omega }_1^2+{\omega }_2^2)g.$
\end{tabular}
\end{equation}
Using these relations  it is not difficult to show that
the quantum dynamical equation

\begin{equation}
\label{fo-q-s}
\frac{\mathrm{d}\hat x_1}{\mathrm{d}t}=\hat x_2,
\quad
\frac{\mathrm{d}\hat x_2}{\mathrm{d}t}=\hat x_3,
\quad
\frac{\mathrm{d}\hat x_3}{\mathrm{d}t}=\hat x_4,
\quad
\frac{\mathrm{d}\hat x_4}{\mathrm{d}t}=
-({\omega }_1^2+{\omega }_2^2)\hat x_3-{\omega }_1^2{\omega }_2^2\hat x_1
\end{equation}
can be represented in the Heisenberg form

\begin{equation}
\label{fo-q-heisenberg}
\mathrm{i}\hbar\frac{\mathrm{d}\hat x_i}{\mathrm{d}t}=
[\hat x_i,\hat H]_{f,g}
\end{equation}
where the quantum Hamiltonian $\hat H$ is obtained from the classical
one (\ref{fo-ham+}) by the replacement of the classical
dynamical variables by the quantum dynamical variables
$x_i \rightarrow \hat x_i,\,i=1, ... ,4.$
Remark, that there is no problems with the ordering of the quantum
variables because all the terms of the form
$\hat x_i\hat x_j$ in $\hat H$ include only commutative operators.
It is easy to check that the quantum analogs of all
the above considered classical integrals of motion
$\hat H_1,\hat H_2,\hat C_1,\hat C_2$
commutate with each other.
As in the classical case, fixing the parameters
$f$ and $g$ one can obtain independent realizations of
the quantum dynamical equations in the Heisenberg form
with different Hamiltonians.
For example, putting
$f=-\frac 1{{\omega}_1^2{\omega }_2^2},\,g=0$
(see (\ref{fo-pi1})) or $f=0,\,g=1,$ (see (\ref{fo-pi2})),
we obtain the equations

\[
\mathrm{i}\hbar\frac{\mathrm{d}\hat x_i}{\mathrm{d}t}=
[\hat x_i,\hat C_1]_1, \quad
\mathrm{i}\hbar\frac{\mathrm{d}\hat x_i}{\mathrm{d}t}=
[\hat x_i,\hat C_2]_2
\]
respectively.
In these equations
for the calculation of the commutator $[.,.]_1$
it is necessary use the first fixed pair $f,\,g,$ and for
the calculation of the commutator $[.,.]_2$
it is necessary use the second one.
In both cases the roles of Hamiltonians play the operators

\begin{equation}
\label{fo-q-int}
\begin{tabular}{l}
$\hat C_1= -\frac 12{\omega}_1^2{\omega }_2^2\hat x_2^2 +
\frac 12({\omega}_1^2 +{\omega }_2^2)\hat x_3^2 +\frac 12\hat x_4^2+
{\omega }_1^2{\omega }_2^2\hat x_1\hat x_3,$ \\[6pt]
$\hat C_2= \frac 12{\omega }_1^2{\omega }_2^2\hat x_1^2 +
\frac 12({\omega }_1^2 +{\omega }_2^2)\hat x_2^2-\frac 12x_3^2 +
\hat x_2\hat x_4$.
\end{tabular}
\end{equation}
Thus, we obtain that the quantum version of the fourth order
Pais-Uhlenbeck oscillator is the bi-Hamiltonian system as well.

In view of the linearity of the quantum dynamical equations
(\ref{fo-q-s}) one can easily writes out their operator
solution

\begin{equation}
\label{fo-q-solution}
\begin{tabular}{l}
$\hat x_1=
e^{-\mathrm{i}{\omega }_1t}a_1
+a_2e^{-\mathrm{i}{\omega }_2t}a_2 + \mbox{h.c.}\,,$
\\[6pt]
$\hat x_2=
-\mathrm{i}{\omega }_1e^{-\mathrm{i}{\omega }_1t}a_1
-\mathrm{i}{\omega }_2e^{-\mathrm{i}{\omega }_2t}a_2
+ \mbox{h.c.}\,,$
\\[8pt]
$\hat x_3=
-{\omega }_1^2e^{-\mathrm{i}{\omega }_1t}a_1
-{\omega }_2^2e^{-\mathrm{i}{\omega }_2t}a_2
+ \mbox{h.c.}\,,$
\\[6pt]
$\hat x_4=
\mathrm{i}{\omega}_1^3e^{-\mathrm{i}{\omega }_1t}a_1
+\mathrm{i}{\omega }_2^3e^{-\mathrm{i}{\omega }_2t}a_2
+ \mbox{h.c.}\,$
\end{tabular}
\end{equation}
where we took into account the self-conjugacy of the dynamical variables
$\hat x_i.$
Using the commutation relations (\ref {fo-q-commut}) we obtain
nonzero commutators among the operators $a_i,\,a_i^+,\,i=1,2$

\begin{equation}
\label{fo-q-commut-creation-annihil}
\lbrack a_1,a_1^{+}]=
\frac{\hbar ({\omega }_2^2f+g)}{2{\omega }_1({\omega }_2^2-{\omega }_1^2)},
\quad
[a_2,a_2^{+}]=
-\frac{\hbar ({\omega }_1^2f+g)}{2{\omega }_2({\omega }_2^2-{\omega
}_1^2)}.
\end{equation}
>From the condition

\begin{equation}
[a_1,a_1^{+}]=[a_2,a_2^{+}]=1,
\end{equation}
imposed usually on commutators of creation and annihilation operators,
we uniquely fix the parameters $f$ and $g$

\begin{equation}
\label{fo-q-fix-commut-creation-annihil}
f=\frac 2\hbar ({\omega }_1+{\omega }_2),\quad
g=-\frac 2\hbar ({\omega }_1^3+{\omega }_2^3).
\end{equation}
Substituting the solution
(\ref{fo-q-solution}) in the expressions
for $\hat H_1,\hat H_2,\hat H,\hat C_1,\hat C_2$
and taking into account the commutation relations
with defined above parameters
(\ref{fo-q-fix-commut-creation-annihil}) we obtain

\begin{equation}
\label{fo-q-const-an-creat-HHH}
\begin{tabular}{l}
$\hat H_1=
\hat C_1+{\omega }_2^2\hat C_2=
2{\omega }_1^2({\omega }_2^2-{\omega }_1^2)^2(a_1^{+}a_1+\frac1{2}),$ \\
$\hat H_2=
\hat C_1+{\omega }_1^2\hat C_2=
2{\omega }_2^2({\omega }_2^2-{\omega }_1^2)^2(a_2^{+}a_2+\frac1{2}),$
\end{tabular}
\end{equation}

\begin{equation}
\label{fo-q-const-an-creat-H}
\hat H=a_1\hat H_1 + a_2\hat H_2 =
\hbar{\omega }_1(a_1^{+}a_1+\frac1{2})+
\hbar{\omega }_2(a_2^{+}a_2+\frac1{2})
\end{equation}
and
\begin{equation}
\label{fo-q-const-an-creat-ÑÑ}
\begin{tabular}{l}
$\hat C_1 = 2({\omega }_2^2-{\omega }_1^2)
\left(-{\omega }_1^4(a_1^{+}a_1+\frac1{2})
+{\omega }_2^4(a_2^{+}a_2+\frac1{2}\right),$ \\
$\hat C_2 = 2({\omega }_2^2-{\omega }_1^2)
\left({\omega }_1^2(a_1^{+}a_1+\frac1{2})
-{\omega }_2^2(a_2^{+}a_2+\frac1{2})\right).$ \\
\end{tabular}
\end{equation}
These formulas show that in the case ${\omega }_1\ne {\omega }_2$
the quantum forth order Pais-Uhlenbeck oscillator
is equivalent to the usual anisotropic harmonic oscillator.
Hence, using the operators $a_i^+,$  $a_i$ one can costruct
the standard Hilbert state space $\mathcal{H}$
with the positive normalized basis vectors
\[
|\psi _{mn}> = \frac{1}{\sqrt{m!n!}}(a_1^+)^m(a_2^+)^n|0>,
\]
and the vacuum vector satisfying the condition
\[ a_1|0> = a_2|0> = 0.\]
All quantum integrals of motion $\hat C_i, \hat H_i $
are diagonal in this basis.

\textbf{4.}
Let us return to the general case of the Pais-Uhlenbeck $2n$-th order
oscillator.  Rewrite the equation (\ref{no}) in the form

\[
\prod _{i=1}^n(\frac{\mathrm{d}^2}{\mathrm{d}t^2}+{\omega }_i^2)x =
\sum _{j=0}^n\sigma ^{n}_j
\frac{\mathrm{d}^{2(n-j)}x}{\mathrm{d}t^{2(n-j)}} =0,
\]
where $\sigma ^{n}_j$ is the $j$-th degree elementary symmetric
polinomials in $n$ variables ${\omega }_i^2,\,i=1, ... ,n$

\[
\sigma ^{n}_j =
\sum _{1\le i_1 < ... < i_j \le n}{\omega }_{i_1}^2{\omega }_{i_2}^2
... {\omega }_{i_j}^2,\quad \sigma ^{n}_0 = 1,
\quad 0\le j\le n.
\]
In these notation the equation (\ref{no}) is equalent
to the following system of $2n$ first order differential equations

\begin{equation}
\label{no-s}
\frac{\mathrm{d}x_i}{\mathrm{d}t} = x_{i+1},\quad i =1, ... ,2n-1, \quad
\frac{\mathrm{d}x_{2n}}{\mathrm{d}t} =
-\sum _{j=1}^n\sigma ^{n}_{j} x_{2(n-j)+1},\quad x_1 = x.
\end{equation}
The system (\ref{no-s}) has $n$ integrals of motion.
In the Pais-Uhlenbeck variables $q_i,\,p_i$

\begin{equation}
\label{no-int-n}
q_i = \sum _{j=0}^{n-1}\sigma ^{n-1}_j(\hat i)x_{2(n-j)-1},\quad
p_i = \sum _{j=0}^{n-1}\sigma ^{n-1}_j(\hat i)x_{2(n-j)},
\quad  i = 1, ... , n,
\end{equation}
where $\sigma ^{n-1}_j(\hat i)$ is the $j$-th degree elementary
symmetric polinomials in $n-1$ variables ${\omega }_k^2,\,k=1, ...
,\hat i, ...  ,n$ (the variable  ${\omega }_i^2$ is omited), these
integrals of motion take the form of the harmonic oscillator energy

\[
 H_i = \frac{1}{2}(p_i^2+{\omega }_i^2q_i^2).
\]
In degenerate cases when some of the frequencies from the set
$\Omega$ coincide, related integrals $H_i$ coincide too.
Hence, as in the case of the forth order oscillator,
the role of a Hamiltonian which generates the dynamical equations
(\ref{no-s}) must play an appropriate linear combination of $H_i.$
We put

\begin{equation}
\label{no-ham-gen}
H =\sum _{i=1}^{n}b_iH_i, \quad
b_i = \frac{1}
{
{\omega }_i\prod_{j=1,{\hat i}}^{n}({\omega }_{i}^2-{\omega }_{j}^2)
},
\end{equation}
where the factor with $j=i$ in the product is omited.
This Hamiltonian together with the Poisson structure $\Pi$
defined by the following nonzero components

\begin{equation}
\label{no-poiss-ham-gen}
\begin{tabular}{l}
$\Pi ^{i,i+1+2j} = (-1)^j\tau_{2i-1+2j},
\quad \tau_{k} = 2\sum _{i=1}^{n} {\omega }_{i}^k,$
\\[8pt]
$i = 1, ... , 2n-1,
\quad 0\le j \le \left[\frac{2n -i-1}{2}\right]$
\end{tabular}
\end{equation}
generates the dynamical equations (\ref{no-s}).
In the formula (\ref{no-poiss-ham-gen}) $[a]$  denote an integral part
of a number $a.$

The quantization of the Pais-Uhlenbeck $2n$-th order oscillator we will
realize by the above scheme. Assume that commutation relations among
the operators $\hat x_i,\,i=1, ... ,2n,$
related to dynamical variables have the quasiclassical
form

\begin{equation}
\label{no-cr}
[\hat x_i,\hat x_j] = \mathrm{i}\hbar\,\Pi ^{ij}.
\end{equation}
Using these relations and the quantum Hamiltonian $\hat H$
obtained from the classical one (\ref{no-ham-gen})
by the substitution $x_i \rightarrow \hat x_i,$
one can check that the quantum version of the dynamical equations
(\ref{no-s})

\[
\frac{\mathrm{d}\hat x_i}{\mathrm{d}t} = \hat x_{i+1},
\quad i =1, ... ,2n-1, \quad
\frac{\mathrm{d}\hat x_{2n}}{\mathrm{d}t} =
-\sum _{j=1}^n\sigma ^{n}_{j} \hat x_{2(n-j)+1}
\]
has the Heisenberg form (\ref{fo-q-heisenberg}).
Using  the operator solution of these equations

\begin{equation}
\label{no-q-solution}
\hat x_1 = \sum _{i=1}^n
e^{-\mathrm{i}{\omega }_it}a_i + \mbox{h.c.},
\quad \hat x_{i} = \frac{\mathrm{d}\hat x_{i-1}}{\mathrm{d}t}, \quad
i = 1, ...,2n,
\end{equation}
and the relations (\ref{no-cr}) we obtain the
commutation relations among the creation and annihilation operators
$a_i,\,a_i^+$

\begin{equation}
\label{no-q-commut-creation-annihil}
\lbrack a_i,a_j^{+}]=\delta _{ij},
\quad  \lbrack a_i,a_j]= \lbrack a_i^{+},a_j^{+}] =0,
\quad i,j=1,...,2n.
\end{equation}
The substitution of the solution of
(\ref{no-q-solution}) in the Hamiltonian $\hat H$ gives us
the Hamiltonian of $n$-dimensional anisotropic oscillator

\begin{equation}
\label{no-ham-creation-annihil}
\hat H = \hbar\sum _{i=1}^n {\omega }_i(a_i^{+}a_i + \frac1{2}).
\end{equation}

\textbf{5.}
Let us conclude these remarks by some notes on the degenerate case
of the fourth order oscillator. As it was pointed out above, under the
condition ${\omega }_1={\omega }_2=\omega$ it is convenient to exploit the
independent integrals (\ref{fo-int}) which take take form

\begin{equation}
\begin{tabular}{l}
$C_{s1}=\frac 12x_4^2+{\omega }^2x_3^2+{\omega }^4x_1x_3-\frac 12{\omega }
^4x_2^2,\quad $ \\
$C_{s2}=-\frac 12x_3^2+{\omega }^2x_2^2+\frac 12{\omega }^4x_1^2+x_2x_4.
\quad $
\end{tabular}
\label{fos-int}
\end{equation}
Substituting ${\omega }_1={\omega }_2=\omega$ in the formulas
(\ref{fo-q-commut}) we obtain the following commutation relations

\begin{equation}
\begin{tabular}{ll}
$[\hat x_1,\hat x_2]=\mathrm{i}\hbar f,$
\quad &
$[\hat x_1,\hat x_4]=\mathrm{i}\hbar g,$
\\[8pt]
$[\hat x_2,\hat x_3]=-\mathrm{i}\hbar g,$
\quad  &
$[\hat x_3,\hat x_4]=-\mathrm{i}\hbar ({\omega }^4f+2{\omega}^2g).$
\end{tabular}
\label{fos-brack}
\end{equation}
Using these relations and the solution of the quantum dynamical
equations (\ref{fo-q-s}) in the degenerates case

\begin{equation}
\label{fo-q-solution-sym}
\begin{tabular}{l}
$\hat x_1=
e^{-\mathrm{i}{\omega}t}a_1+te^{-\mathrm{i}{\omega }t}a_2
+ \mbox{h.c.}\,,$
\\[8pt]
$\hat x_2=
-\mathrm{i}{\omega }e^{-\mathrm{i}{\omega }t}a_1+
(1-\mathrm{i}{\omega}t)e^{-\mathrm{i}{\omega }t}a_2 + \mbox{h.c.}\,,$
\\[12pt]
$\hat x_3=
-{\omega }^2e^{-\mathrm{i}{\omega }_1t}a_1
-{\omega }(2\mathrm{i}+{\omega }t)e^{-\mathrm{i}{\omega }t}a_2
+ \mbox{h.c.}\,,$
\\[8pt]
$\hat x_4=
\mathrm{i}{\omega}^3e^{-\mathrm{i}{\omega }_1t}a_1+
{\omega }^2(-3+\mathrm{i}t{\omega })e^{-\mathrm{i}{\omega }t}a_2
+ \mbox{h.c.}$
\end{tabular}
\end{equation}
we obtain the commutation relations among the creation and annihilaton
operators $a_i,a_i^+$

\begin{equation}
\label{}
[a_1,a_1^+] =\frac{\hbar}{4}\frac{3{\omega}^2f+g}{{\omega}^3},
\quad
[a_2,a_2^+] = 0,
\quad
[a_1,a_2^+] = [a_1^+,a_2] =
-\frac{\mathrm{i}\hbar}{4}\frac{{\omega}^2f+g}{{\omega}^2}.
\end{equation}
Remark that the pair $a_2,a_2^+$ commutes for any parameters
$f,g$. It is useful to fix these parameters by the conditions

\[
[a_1,a_1^+]=1,\quad [a_2,a_2^+]=[a_1,a_2^+]=[a_2,a_1^+]= 0.
\]
We obtain
$g=-\omega ^2f$  and $f=\frac{2\omega}{\hbar}.$
The integrals of motions (\ref{fos-int}) in the terms of $a_i^+,a_i$
have the form

\[
C_{s1} = 16{\omega}^4a_2^+a_2 +
4\mathrm{i}{\omega}^5(a_2a_1^+-a_1a_2^+),
\]
\[
C_{s2} = -8{\omega}^4a_2^+a_2 -
4\mathrm{i}{\omega}^3(a_2a_1^+-a_1a_2^+).
\]
Commutativity of the operators $a_2,a_2^+$ tells us that
constructing the state space of this
Pais-Uhlenbeck oscillator it is necessary to take into account
their classical character.

\vspace{0.5cm}
\begin{center} {\Large\bf Acknowledgments} \end{center}
\vspace{0.3cm}
The authors would like to thank P.P.Kulish for useful
discussions and references.
This work was supported in part by the Russian Fond of Basic
Researches (grant No 06-01-00451).

\end{document}